\documentclass{article}
\usepackage{ismir,amsmath,cite}
\usepackage{aliascnt}
\usepackage{tabu}
\usepackage{amssymb,amsmath,mathtools}
\usepackage{booktabs}
\usepackage{multirow}
\usepackage[english]{babel}

\usepackage{times}
\usepackage{url}
\usepackage{color}
\usepackage{multirow}
\usepackage{multicol}
\usepackage{graphicx}
\usepackage{caption}
\usepackage{tikz}
\usetikzlibrary{calc,tikzmark}
\usepackage[shortcuts]{extdash}
\usepackage{blindtext}
\usepackage[ruled,norelsize,linesnumbered]{algorithm2e}

\newcommand{\comment}[1]{\textcolor{blue}{}}
\newcommand{\mg}[1]{\textcolor{red}{}}

\makeatletter
\def\inmod#1{\allowbreak\mkern5mu{\operator@font mod}\,\,#1}
\makeatother

\urldef{\discovery}\url{http://www.music-ir.org/mirex/wiki/2017:Discovery_of_Repeated_Themes_%26_Sections}

\title{Learning transposition-invariant interval features from symbolic music and audio}
\multauthor
{Stefan Lattner$^{1,2}$, Maarten Grachten$^{1,2}$, Gerhard Widmer$^1$}
{$^1$ Institute of Computational Perception, JKU Linz\\
$^2$ Sony Computer Science Laboratories (CSL), Paris, France}
\def\authorname{Stefan Lattner, Maarten Grachten, Gerhard Widmer}

\begin{document}

\maketitle

\begin{abstract}
Many music theoretical constructs (such as scale types, modes, cadences, and chord types) are defined in terms of pitch intervals---relative distances between pitches. 
Therefore, when computer models are employed in music tasks, it can be useful to operate on interval representations rather than on the raw musical surface.
Moreover, interval representations are transposition-invariant, valuable for tasks like audio alignment, cover song detection and music structure analysis.
We employ a gated autoencoder to learn fixed-length, invertible and transposition-invariant interval representations from polyphonic music in the symbolic domain and in audio.
An unsupervised training method is proposed yielding an organization of intervals in the representation space which is musically plausible.
Based on the representations, a transposition-invariant self-similarity matrix is constructed and used to determine repeated sections in symbolic music and in audio, yielding competitive results in the MIREX task "Discovery of Repeated Themes and Sections".
\end{abstract}

\section{Introduction}\label{sec:introduction}
The notion of relative pitch is important in music understanding. Many music theoretical concepts, such as scale types, modes, chord types and cadences, are defined in terms of relations between pitches or pitch classes.
But relative pitch is not only a music theoretical construct.
It is common for people to perceive and memorize melodies in terms of pitch intervals (or in terms of \emph{contours}, the upward or downward \emph{direction} of pitch intervals) rather than sequences of absolute pitches.
This characteristic of music perception also has ramifications for the perception of form in musical works, since it implies that transposition of some musical fragment along the pitch dimension (such that the relative distances between pitches remain the same) does not alter the perceived identity of the musical material, or at least establishes a sense of similarity between the original and the transposed material.
As such, adequate detection of musical form in terms of (approximately) repeated structures presupposes the ability to account for pitch transposition---one of the most common types of transformations found in music.

Relative pitch perception in humans is currently not well-understood~\cite{mcdermott08d}. For example there are no established theories on how the human brain derives a relative representation of pitch from the tonotopic representations formed in the cochlea, neither is it clear whether there is a connection between the perception of pitch relations in simultaneous versus consecutive pitches.

Computational approaches to address tasks of music understanding (such as detecting patterns and form in music) often circumvent this issue by representing musical stimuli as sequences of monophonic pitches, after which simply differencing consecutive pitches yields a relative pitch representation.
This approach also works for polyphonic music, to the extent that the music can be meaningfully segregated into monophonic pitch streams.
A drawback of this approach is that it presupposes the ability to segregate musical streams, which is often far from trivial due to the ambiguity of musical contexts.
To take an analogous approach on acoustical representations of musical stimuli is even more challenging, since it further depends on the ability to detect pitches and onsets in sound.

In this paper we take a different approach altogether.
We train a neural network model to learn representations that represent the relation between the music at some time point $t$ and the preceding musical context.
During training, these representations are adapted to minimize the reconstruction error of the music at $t$ given the preceding context and the representation itself.

A crucial aspect of the model is its bilinear architecture (more specifically, a \emph{gated autoencoder}, or GAE architecture) involving multiplicative connections, which facilitates the formation of relative pitch representations.
We stimulate such representations more explicitly using an altered training procedure in which we transpose the training data using arbitrary transpositions.

The result are two models (for symbolic music and audio) that can map both monophonic and polyphonic music to a sequence of points in a vector space---the \emph{mapping space}---in a way that is invariant to pitch transpositions.
This means that a musical fragment will be projected to the same mapping space trajectory independently of how it is transposed.

We validate our approach experimentally in several ways. First we show that musical fragments that are nearest neighbors in the mapping space have many pitch intervals in common (as opposed to nearest neighbors in the input space). Then we show that the topology of the learned mapping space reflects musically meaningful relations between intervals (such as the tritone being dissimilar to other intervals). Lastly we use mapping space representations to detect musical form both for symbolic and audio representations of music, showing that it yields competitive results, and in the case of audio even improves the state of the art.
A re-implementation of the transposition-invariant GAE for audio is publicly available\footnote{see \url{https://github.com/SonyCSLParis/cgae-invar}}.

The paper is structured as follows.
Section~\ref{sec:related_work} provides an overview of relation learning using GAEs, and reviews work on creating interval representations from music.
In Section~\ref{sec:model}, the used architecture is described and in Section~\ref{sec:data}, data is introduced on which the GAE is trained.
The training procedure, including the novel method to support the emergence of transposition-invariance, is proposed in Section~\ref{sec:training}.
The experiments conducted to examine the properties of learned mappings are described in Section~\ref{sec:experiments}, and results are presented and discussed in Section~\ref{sec:res}.
Section~\ref{sec:concl-future-work} wraps the paper up with conclusions and prospects of future work.

\section{Related Work}\label{sec:related_work}
GAEs utilize \emph{multiplicative interactions} to learn correlations between or within data instances.
The method was inspired by the correlation theory of the brain \cite{von1994correlation}, where it was pointed out that some cognitive phenomena cannot be explained with the conventional brain theory and an extension was proposed which involves the correlation of neural patterns.

In machine learning, this principle was deployed in \emph{bi-linear} models, for example to separate person and pose in face images \cite{tenenbaum2000separating}.
Bi-linear models, like the GAE, are two-factor models whose outputs are linear in either factor when the other is held constant.
\cite{olshausen2007bilinear} proposed another variant of a bi-linear model in order to learn objects and their optical flow.
Due to its similar architecture, the gated Boltzmann machine (GBM) \cite{memisevic2007unsupervised, memisevic2010learning} can be seen as a direct predecessor of the GAE.
The GAE was introduced by \cite{memisevic2011gradient} as a derivative of the GBM, as standard learning criteria became applicable through the development of denoising autoencoders \cite{vincent2010stacked}.

GAEs have been further used to learn transformation-invariant representations for classification tasks \cite{ICML2012Memisevic_105}, for parent-offspring resemblance \cite{dehghan2014look}, for learning to negate adjectives in linguistics \cite{rimell2017learning}, for activity recognition with the Kinekt sensor \cite{mocanu2015factored}, in robotics to learn to write numbers \cite{droniou2014learning}, and for learning multi-modal mappings between action, sound, and visual stimuli \cite{droniou2015deep}.


In music, bi-linear models have been applied to learn co-variances within spectrogram data for music similarity estimation \cite{schluter2011music}, and for learning musical transformations in the symbolic domain \cite{lattner2017relations}.
In sequence modeling, the GAE has been utilized to learn co-variances between subsequent frames in movies of rotated 3D objects \cite{memisevic2013aperture} and to predict accelerated motion by stacking more layers in order to learn higher-order derivatives \cite{michalski2014modeling}, which uses a method similar to the one proposed here.

Transposition-invariance in music is achieved in \cite{meredith2002algorithms} by transforming symbolic pitch--time representations into point-sets, in which translatable patterns are identified.
Another method in the symbolic domain is that in \cite{cambouropoulos1996general}, where a general interval representation for polyphonic music is put forward, in  \cite{nakamura2015characteristics}, where specific pitch-class intervals in polyphonic music are used for characterizing music styles and in \cite{DBLP:conf/ismir/MullerC07} where transposition-invariant self-similarity matrices are computed.
In \cite{marolt2008mid}, an approach to calculating transposition-invariant mid-level representations from audio is introduced, based on the 2-D power spectrum of melodic fragments.
Similarly, a method to calculate interpretable interval representations from audio is proposed in \cite{walters2012intervalgram}, where chromagrams that are close in time are cross-correlated to obtain local pitch-invariance.

\section{Model}\label{sec:model}
\begin{figure}
\begin{center}
\includegraphics[width=.8\linewidth]{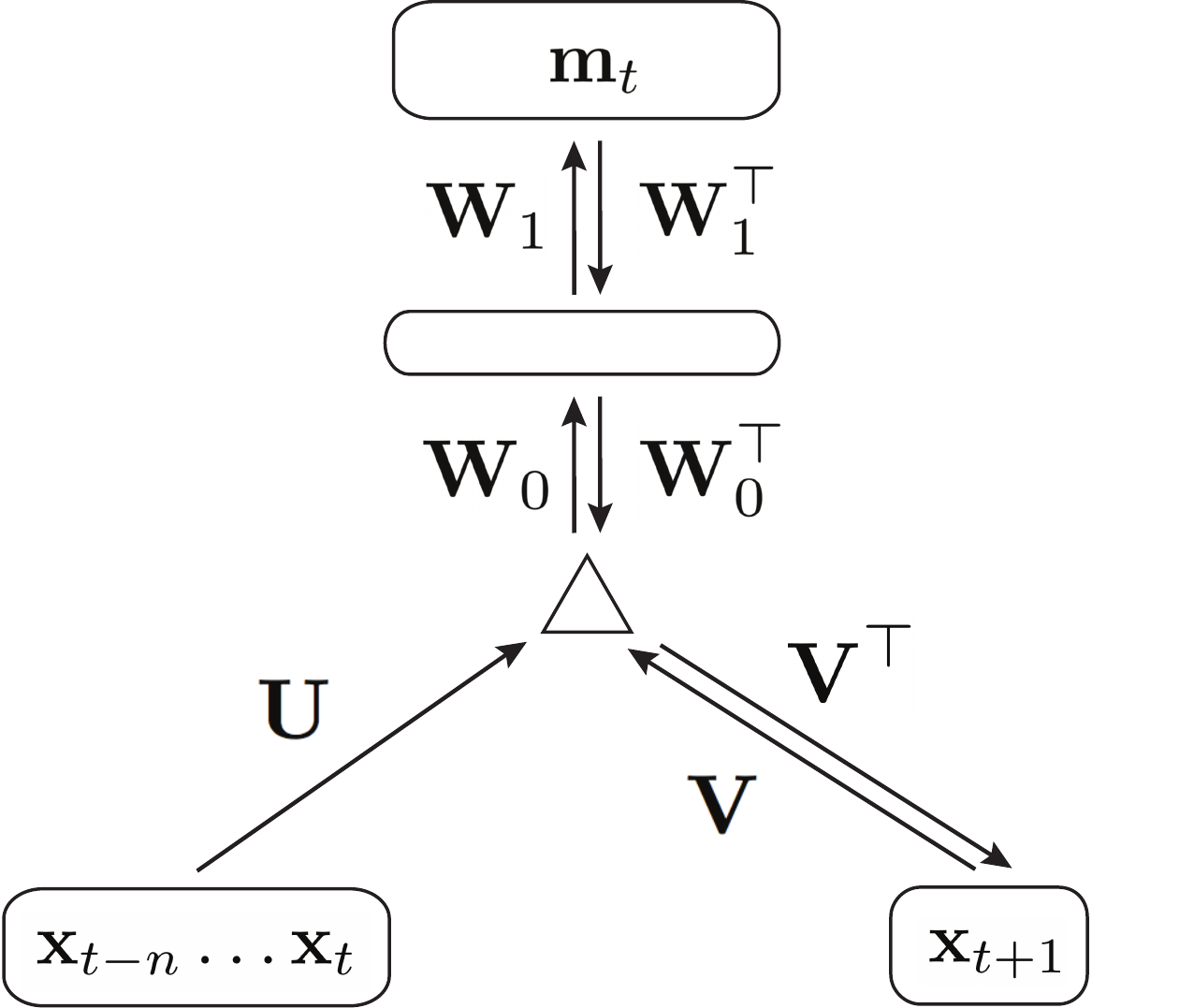}
\caption{Schematic illustration of the gated autoencoder architecture used in the experiments.}
\label{fig:ga}
\end{center}
\end{figure}

Let $\mathbf{x}_j$ be a vector representing pitches of currently sounding notes (in the symbolic domain) or the energy distributed over frequency bands (in the audio domain), in a fixed-length time interval.
Given a temporal context $\mathbf{x}_{t-n}^{t} = \mathbf{x}_{t-n} \dots \mathbf{x}_{t}$ as the input and the next time step $\mathbf{x}_{t+1}$ as the target, the goal is to learn a mapping $\mathbf{m}_t$ which does not change when shifting $\mathbf{x}_{t-n}^{t+1}$ up- or downwards in the pitch dimension.
A gated autoencoder (GAE, depicted in Figure~\ref{fig:ga}) is well-suited for this task,  modeling the intervals between reference pitches in the input and pitches in the target, encoded in the latent variables of the GAE as mapping codes $\mathbf{m}_j$.
Unlike in common prediction tasks, the targets are known when training a GAE.
The goal of the training is to find a mapping $\mathbf{m}_j$ for any input/target pair which transforms the input into the given target.
The mapping at time $t$ is calculated as

\begin{equation}\label{eq:gamap}
\mathbf{m}_t = \sigma_h (\mathbf{W}_1 \sigma_h(\mathbf{W}_0(\mathbf{U}\mathbf{x}_{t-n}^{t} \cdot \mathbf{V}\mathbf{x}_{t+1}))),
\end{equation}
where $\mathbf{U}, \mathbf{V}$ and $\mathbf{W}_k$ are weight matrices, $\sigma_h$ is the hyperbolic tangent non-linearity, and we will refer to the learnt mappings $\mathbf{m}_j$ as the \emph{mapping space} of the input/target pairs. 
The operator~$\cdot$ (depicted as a triangle in Figure~\ref{fig:ga}) depicts the Hadamard (or element-wise) product of the filter responses $\mathbf{U}\mathbf{x}_{t-n}^{t}$ and $\mathbf{V}\mathbf{x}_{t+1}$, denoted as \emph{factors}.
This operation allows the model to \emph{relate} its inputs, making it possible to learn interval representations.

The target of the GAE can be reconstructed as a function of the input $\mathbf{x}_{t-n}^{t}$ and a mapping $\mathbf{m}_t$:
\begin{equation}\label{recony}
\mathbf{\tilde{x}}_{t+1} = \sigma_g(\mathbf{V}^\top (\mathbf{W}_0^\top\mathbf{W}_1^\top \mathbf{m}_t \cdot \mathbf{U}\mathbf{x}_{t-n}^{t})),
\end{equation}
where $\sigma_g$ is the sigmoid non-linearity for binary input and the identity function for real-valued input.

The cost function is defined to penalize the error of reconstructing the target $\mathbf{x}_{t+1}$ given the input $\mathbf{x}_{t-n}^{t}$ and the mapping $\mathbf{m}_t$ as
\begin{equation}\label{eq:cost}
\mathcal{L}_{c} = c(\mathbf{x}_{t+1}, \mathbf{\tilde{x}}_{t+1}),
\end{equation}
where $c(\cdot)$ is the mean-square error for real-valued sequen-ces and the cross-entropy loss for binary sequences.

\section{Data}\label{sec:data}
We train the model both on symbolic music representations and on audio spectrograms.
For the symbolic data, the Mozart/Batik data set~\cite{widmer2003discovering} is used, consisting of 13 piano sonatas containing more than 106,000 notes.
The dataset is encoded as successive $60$ dimensional binary vectors (encoding MIDI note number $36$ to $96$), each representing a single time step of 1/16th note duration.
The pitch of an active note is encoded as a corresponding on-bit, and as multiple voices are encoded simultaneously, a vector may have multiple active bits.
The result is a pianoroll-like representation.

The audio dataset consists of 100 random piano pieces of the MAPS dataset \cite{emiya2010multipitch} (subset MUS), at a sampling rate of 22.05 kHz. We choose a constant-Q transformed spectrogram using a hop size of $1984$, and Hann windows with different sizes depending on the frequency bin. The range comprises $120$ frequency bins (24 per octave), starting from a minimal frequency of $65.4$ Hz. Each time step is contrast-normalized to zero mean and unit variance.

\section{Training}\label{sec:training}
The model is trained with stochastic gradient descent in order to minimize the cost function (cf. Equation~\ref{eq:cost}) using the data described in Section~\ref{sec:data}.
However, rather than using the data as is, we use data-augmentation in combination with an altered training procedure to explicitly aim at transposition invariance of the mapping codes.

\subsection{Enforcing Transposition-Invariance}\label{sec:enforc-transp-invar}
As described in Section~\ref{sec:model} the classical GAE training procedure derives a mapping code from an input/target pair, and subsequently penalizes the reconstruction error of the target given the input and the derived mapping code.
Although this procedure naturally tends to lead to similar mapping codes for input target pairs that have the same interval relationships, the training does not explicitly enforce such similarities and consequently the mappings may not be maximally transposition invariant.

Under ideal transposition invariance, by definition the mappings would be identical across different pitch \linebreak transpositions of an input/target pair. Suppose that a pair $(\mathbf{x}_{t-n}^{t}, \mathbf{x}_{t+1})$ leads to a mapping $\mathbf{m}$ (by Equation~\ref{eq:gamap}).
Transposition invariance implies that reconstructing a target $\mathbf{x}^{\prime}_{t+1}$ from the pair $({\mathbf{x}^{\prime}}_{t-n}^{t}, \mathbf{m})$ should be as successful as reconstructing $\mathbf{x}_{t+1}$ from the pair $(\mathbf{x}_{t-n}^{t}, \mathbf{m})$ when $({\mathbf{x}^{\prime}}_{t-n}^{t}, \mathbf{x}^{\prime}_{t+1})$ can be obtained from $(\mathbf{x}_{t-n}^{t}, \mathbf{x}_{t+1})$ by a single pitch transposition.

Our altered training procedure explicitly aims to achieve this characteristic of the mapping codes by penalizing the reconstruction error using mappings obtained from transposed input/target pairs.
More formally, we define a transposition function $\textit{shift}(\mathbf{x}, \delta)$, shifting the values of a vector $\mathbf{x}$ of length $M$ by $\delta$ steps (MIDI note numbers and CQT frequency bins for symbolic and audio data, respectively):
\begin{equation}\label{eq:shift}
\textit{shift}(\mathbf{x}, \delta) = (x_{(0+\delta)\inmod{M}}, \dots, x_{(M-1+\delta)\inmod{M}})^\top,
\end{equation}
and $\textit{shift}(\mathbf{x}_{t-n}^{t}, \delta)$ denotes the transposition of each single time step vector \emph{before} concatenation and linearization.


The training procedure is then as follows. First, the mapping code $\mathbf{m}_{t}$ of an input/target pair is inferred as shown in Equation \ref{eq:gamap}. Then, $\mathbf{m}_{t}$ is used to reconstruct a \emph{transposed} version of the target, from an equally \emph{transposed} input (modifying Equation \ref{recony}) as
\begin{equation}\label{reconyshift}
\mathbf{\tilde{x}}'_{t+1} = \sigma_g(\mathbf{V}^\top (\mathbf{W}_0^\top\mathbf{W}_1^\top \mathbf{m}_t \cdot \mathbf{U}\textit{shift}(\mathbf{x}_{t-n}^{t},\delta))),
\end{equation}
with $\delta \in [-30,30]$ for the symbolic, and  $\delta \in [-60,60]$ for the audio data.
Finally, we penalize the error between the reconstruction of the transposed target and the actual transposed target (i.e., employing Equation \ref{eq:cost}) as
\begin{equation}
\mathcal{L}(\textit{shift}(\mathbf{x}_{t+1},\delta),\mathbf{\tilde{x}}'_{t+1}).
\end{equation}

\begin{figure}
\includegraphics[width=1.\linewidth]{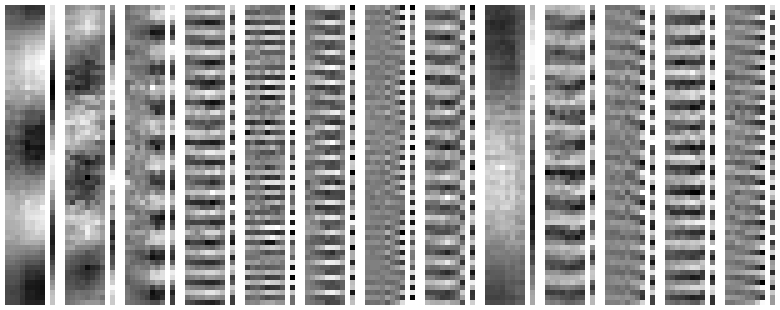} \\
\vspace{-8mm} \\
\caption{Some filter pairs $\in$ \{$\mathbf{U}, \mathbf{V}$\} of a GAE trained on polyphonic Mozart piano pieces.
}
\label{fig:filters}
\end{figure}

The transposition distance $\delta$ is randomly chosen for each training batch.
This method amounts to both, a form of guided training and data augmentation.
Some weights (i.e., filters) in $\mathbf{U}$ and $\mathbf{V}$ resulting from that training are depicted in Figure \ref{fig:filters}.

\subsection{Architecture and Training Details}
The architecture and training details of the GAE are as follows: A temporal context length of $n = 8$ is used (the choice of $n>1$ leads to higher robustness of the mapping codes to diatonic transposition). The factor layer has $1024$ units for the symbolic data, and $512$ units for the spectrogram data.
Furthermore, for all datasets, there are $128$ neurons in the first mapping layer and $64$ neurons in the second mapping layer (resulting in $\mathbf{m}_t \in \mathbb{R}^{64}$). 

L2 weight regularization for weights $\mathbf{U}$ and $\mathbf{V}$ is applied, as well as sparsity regularization \cite{Lee:2007uz} on the topmost mapping layer. 
The deviation of the norms of the columns of both weight matrices $\mathbf{U}$ and $\mathbf{V}$ from their average norm is penalized. 
Furthermore, we restrict these norms to a maximum value.
We apply $50\%$ dropout on the input and no dropout on the target, as proposed in \cite{memisevic2011gradient}. 
The learning rate (1e-3) is gradually decremented to zero over the course of training.


\section{Experiments}\label{sec:experiments}
In this Section we describe several experimental analyses to validate the proposed approach.
They are intended to test the degree of transposition-invariance of the learned mappings, as well as assess their musical relevance (Sections~\ref{sec:class-clust-analys} and~\ref{sec:sensitivity-analysis}). Finally, we put the learned representations to practice in a repeated section discovery task for symbolic music and audio (Section~\ref{sec:disc-repe-them}).

\subsection{Classification and Cluster Analysis}\label{sec:class-clust-analys}
Our hypothesis is that the model learns relative pitch representations (i.e. intervals) from polyphonic absolute pitch sequences.
In order to test this hypothesis, we conduct two experiments using the symbolic data.

In the first experiment a ten-fold k-nn classification of intervals is performed (where k = 10), where the task is to identify all pitch intervals between notes in the input and the target of an input/target pair.
If the learned mappings actually represent intervals, the classifier will perform substantially better on the mappings than on the input space.
As intervals in music are transposition-invariant, the interval labels do not change when performing transposition in the input space.
Thus, we perform the classification on the mappings of the original data and of randomly transposed data, to test if the mappings are indeed transposition-invariant.

We label the symbolic train data input/target pairs according to all intervals which occur between them, independent of the temporal distance of the notes exhibiting the intervals. Thus, each pair can have multiple labels.
For each pair in the test set the k-nn classifier predicts the set of interval labels that are present in the k neighbors of that pair.
The classification is performed in the input space (using concatenated pairs) and in the mapping space.
Using these predictions we determine the precision, recall, and F-score over the test set (cf. Table~\ref{tab:knn}).
For example, when a pair contains 6 intervals and the classifier estimate yield 4 true-positive and 4 false-positive interval occurrences, that pair is assigned a precision of 0.5 and a recall of 0.67.

\begin{figure}
\begin{center}
\includegraphics[width=1.\linewidth]{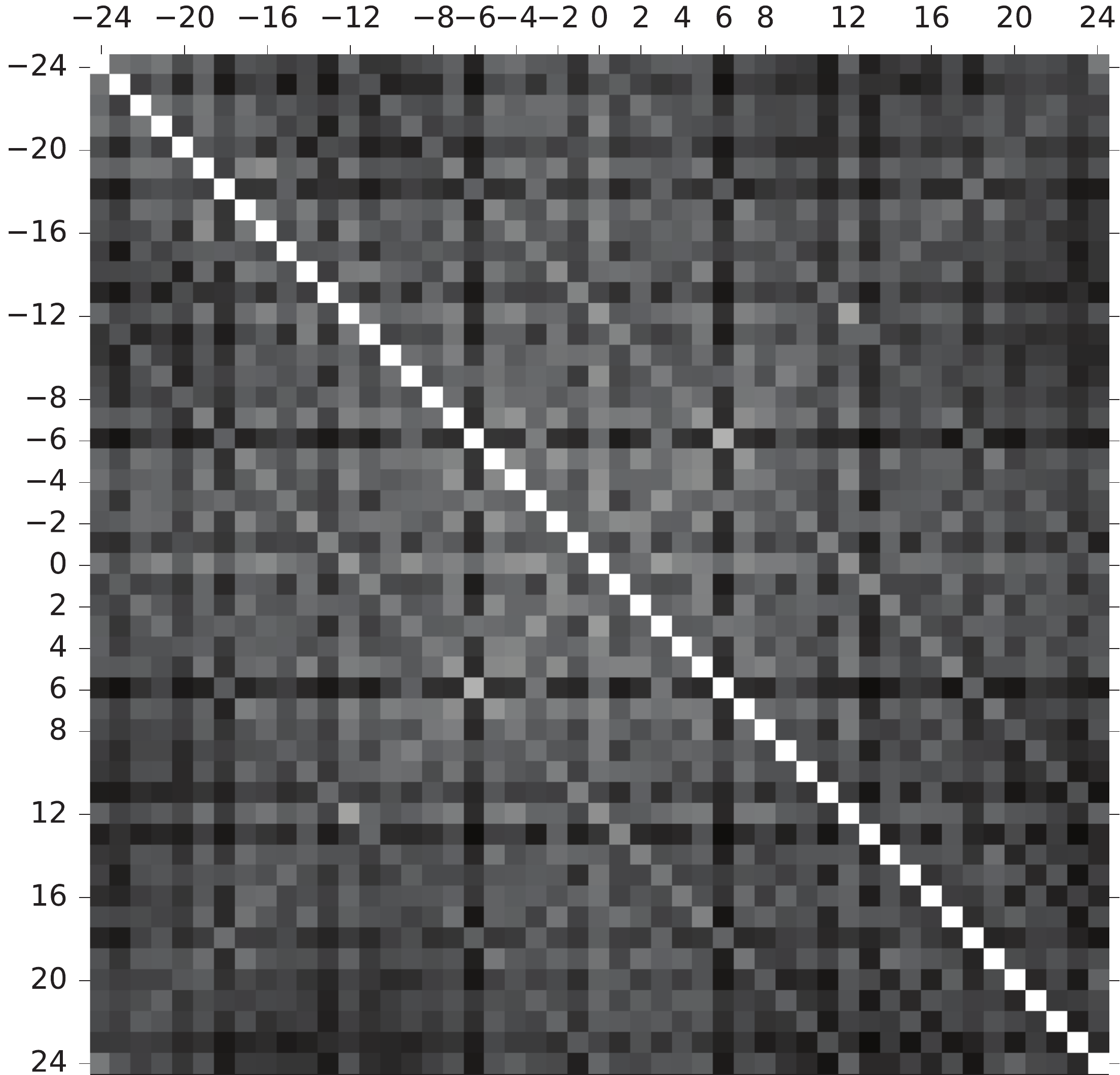}
\caption{Distance matrix of cluster centers of intervals represented in mapping space. Darker cells indicate higher distances between respective clusters, brighter cells indicate closeness.}
\label{fig:close_matrix}
\end{center}
\end{figure}

In the second part of the experiment, the cluster centers of all intervals in the mapping space are determined.
Again, each pair projected into the mapping space accounts for all intervals it exhibits and can therefore participate in more than one cluster.
The mutual Euclidean distances between all cluster centers are displayed as a matrix (cf. Figure~\ref{fig:close_matrix}).
An interpretation of the results follows in Section~\ref{sec:res}.

\begin{table}[t]
\centering
\begin{tabular}{llll}
\toprule
Data     & Precision & Recall & F1 \\
\midrule
\rule{-6pt}{0ex}
\textbf{Original input} & & & \\
\rule{-2pt}{2ex}
Mapping space & \textbf{91.27}     & 70.25  & 76.66   \\
Input space & 65.58    & 46.05  & 50.59   \\
\midrule
\rule{-6pt}{0ex}
\textbf{Transposed input} & & & \\
\rule{-2pt}{2ex}
Mapping space & 90.78     & \textbf{71.44}  & \textbf{77.31}   \\
Input space & 51.81    & 32.99  & 37.43   \\
\midrule
All & 26.40     & 100.0   & 40.05  \\
None & 0.0 & 0.0 & 0.0 \\
\bottomrule
\end{tabular}
\caption{Results of the k-nn classification in the mapping space and in the input space for the original symbolic data and data randomly transposed by $[-24,24]$ semitones. ``All'' is a lower bound (always predict all intervals), ``None'' returns the empty set.}
\label{tab:knn}
\end{table}

\subsection{Discovery of Repeated Themes and Sections}\label{sec:disc-repe-them}
\urldef\mirexurl\url{http://www.music-ir.org/mirex/wiki/2017:Discovery_of_Repeated_Themes_&_Sections}
The MIREX Task for Discovery of Repeated Themes and Sections for Symbolic Music and Audio\footnote{\mirexurl{}} 
tests algorithms for their ability to identify repeated patterns in music.
The commonly used JKUPDD dataset \cite{collins2017discovery} contains 26 motifs, themes, and repeated sections annotated in 5 pieces by J. S. Bach, L.~v.~Beethoven, F. Chopin, O. Gibbons and W. A. Mozart.
We use the MIDI and the audio versions of the dataset and preprocess them as described in Section~\ref{sec:data}.

\begin{figure}[t]
\begin{center}
\includegraphics[trim=0cm 1cm 0cm 1cm, clip=true, width=1.\linewidth]{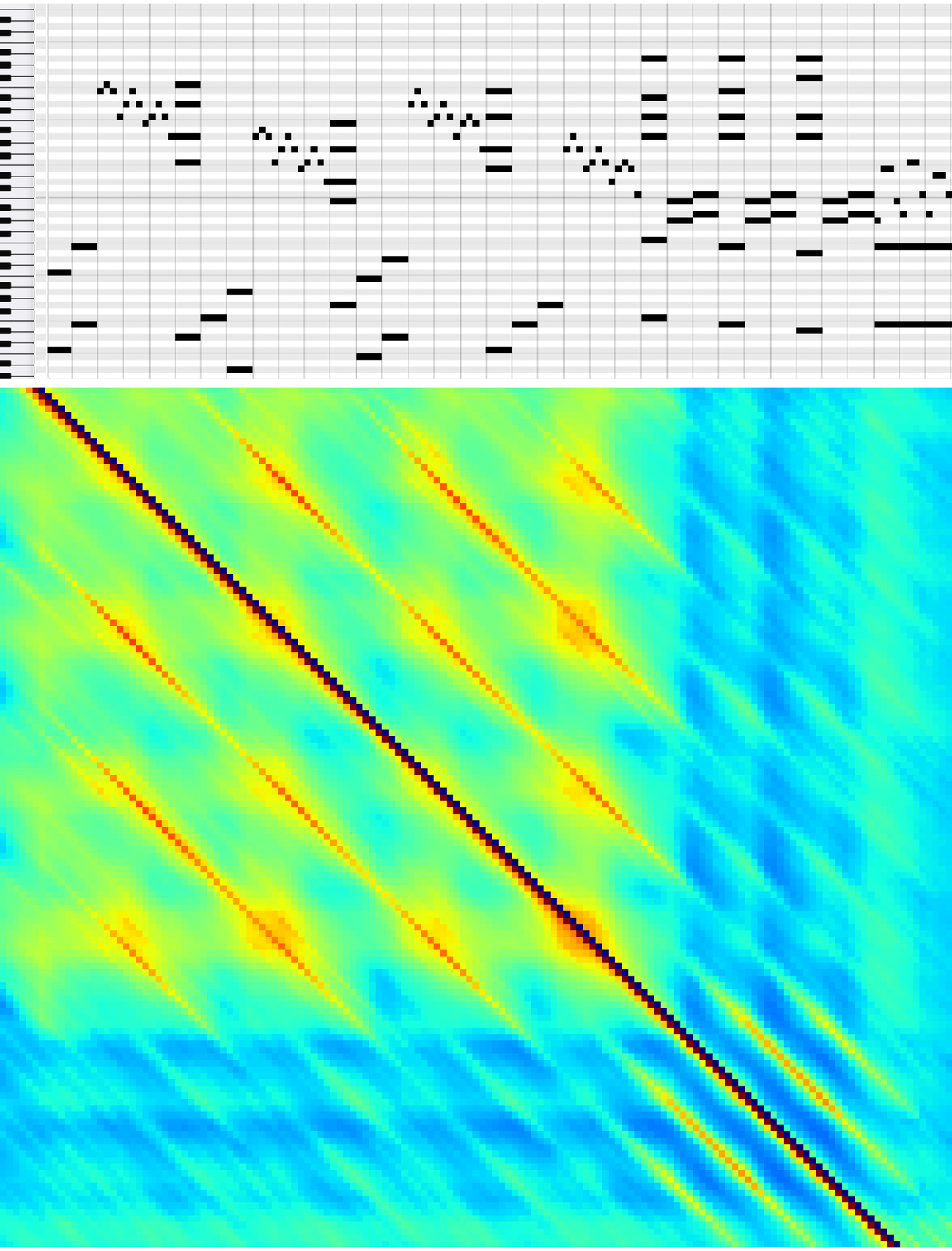}
\caption{Symbolic music and corresponding self-similarity matrix calculated from transposition-invariant mapping codes. Warmer colors indicate similarity, colder colors indicate dissimilarity.}
\label{fig:self-sim-invar}
\end{center}
\end{figure}

We calculate the reciprocal of the Euclidean distances between all representations $\mathbf{m}_t$ of a song, resulting in a transposition-invariant similarity matrix $X$.
Then, the values of the main diagonal are set to the minimal value of the matrix.
Subsequently, the matrix is normalized and convolved with an identity matrix of size $15 \times 15$ to emphasize and smooth diagonals (Figure~\ref{fig:self-sim-invar} shows a resulting matrix). 
The method used to determine repeated parts based on diagonals of high values in the self-similarity matrix is adopted from \cite{nieto2014identifying}, with a different method to identify diagonals, as described below.

The function 
\begin{equation}\label{eq:diag1}
s(i,j,N) = \sum_{k=N-m}^{N}{\frac{X(i+k,j+k) w_k}{m} }
\end{equation}
returns the score for a diagonal starting at $X(i,j)$ with length $N$, and diagonals with high score are considered to be repeated sections. For each $i,j$, we iteratively evaluate the score with $N$ increasing from $1$ in integer steps, until the score undercuts a threshold $\gamma$. Only the last $m$ values, $m = \min(10,N)$, of the diagonal are taken into account, because those values indicate when to stop tracing.
The factor
\begin{equation}\label{eq:diag2}
w_k = \frac{1+k+m-N}{m}
\end{equation}
linearly weights the last $m$ values of the diagonal so that later values have more impact on the overall score.

Three empirically determined parameters influence the functioning of the method: (1) from the diagonals found, we only keep those spanning more than $\mathit{2}$ \emph{whole notes}, (2) all sections whose common boundaries start and end within the length of \emph{a half note} are considered to be repetitions of each other, (3) the thresholds $\gamma$ determining if a diagonal should be considered a repetition in the symbolic and the audio data are set to $0.9$ and $0.81$, respectively. The results are shown in Table~\ref{tab:secdiscovery} and are discussed in Section~\ref{sec:res}.

\begin{table*}
\centering
\footnotesize
\begin{tabular}{llllllllllllll}
\toprule
Algorithm     & $F_{\text{est}}$  & $P_{\text{est}}$  & $R_{\text{est}}$  & $F_{\text{o(.5)}}$   & $P_{\text{o(.5)}}$   & $R_{\text{o(.5)}}$
 & $F_{\text{o(.75)}}$   & $P_{\text{o(.75)}}$   & $R_{\text{o(.75)}}$  & $\mathbf{F_3}$    & $P_3$    & $R_3$    & Time (s) \\
\midrule
\rule{-6pt}{0ex}
\textbf{Symbolic} & & & & & & & & & & & & & \\
\rule{-2pt}{2ex}
GAE intervals (ours) & 59.07 & \textbf{77.60} & 58.30 & 68.92 & \textbf{80.24} & 67.46 & \textbf{77.51} & \textbf{91.38} & 73.29 & 50.44 & 60.36 & 53.23 & \textbf{127}  \\
VMO symbolic \cite{wang2015music}  & \textbf{60.79} & 74.57 & 56.94 & 71.92 & 79.54 & 68.78 & 75.98 & 75.98 & 75.99 & \textbf{56.68} & \textbf{68.98} & 53.56 & 4333 \\
SIARCT-CFP \cite{collins2013siarct}  & 33.70  & 21.50  & \textbf{78.00}  & \textbf{76.50}  & 78.30  & \textbf{74.70}  & -     & -     & -     & -     & -     & -     & -    \\
COSIATEC \cite{meredith2013cosiatec}  & 50.20  & 43.60  & 63.80  & 63.20  & 57.00  & 71.60  & 68.40  & 65.40  & \textbf{76.40}  & 44.20  & 40.40  & \textbf{54.40}  & 7297 \\
\midrule
\rule{-6pt}{0ex}
\textbf{Audio} & & & & & & & & & & & & & \\
\rule{-2pt}{2ex}
GAE intervals (ours) & \textbf{57.67} & \textbf{67.46} & 59.52 & 58.85 & 61.89 & 56.54 & 68.44 & 72.62 & 64.86 & \textbf{51.61} & 59.60 & \textbf{55.13}  & 194 \\ 
VMO deadpan \cite{wang2015music} & 56.15 & 66.80  & 57.83 & \textbf{67.78} & \textbf{72.93} & \textbf{64.30}  & \textbf{70.58} & \textbf{72.81} & \textbf{68.66} & 50.60  & \textbf{61.36} & 52.25 & \textbf{96}   \\
SIARCT-CFP \cite{collins2013siarct}    & 23.94 & 14.90  & \textbf{60.90}  & 56.87 & 62.90 & 51.90  & -     & -     & -     & -     & -     & -     & -    \\
Nieto  \cite{nieto2014identifying}      & 49.80  & 54.96 & 51.73 & 38.73 & 34.98 & 45.17 & 31.79 & 37.58 & 27.61 & 32.01 & 35.12 & 35.28 & 454 \\
\bottomrule
\end{tabular}
\caption{Different precision, recall and f-scores (adopted from \cite{wang2015music}, details on the measures are given in \cite{collins2017discovery}) of different methods in the Discovery of Repeated Themes and Sections MIREX task, for symbolic music and audio. The $\mathbf{F_3}$ score constitutes a summarization of all measures.}
\label{tab:secdiscovery}
\end{table*}

\subsection{Sensitivity Analysis}\label{sec:sensitivity-analysis}
\begin{figure}
\begin{center}
\includegraphics[width=1.\linewidth]{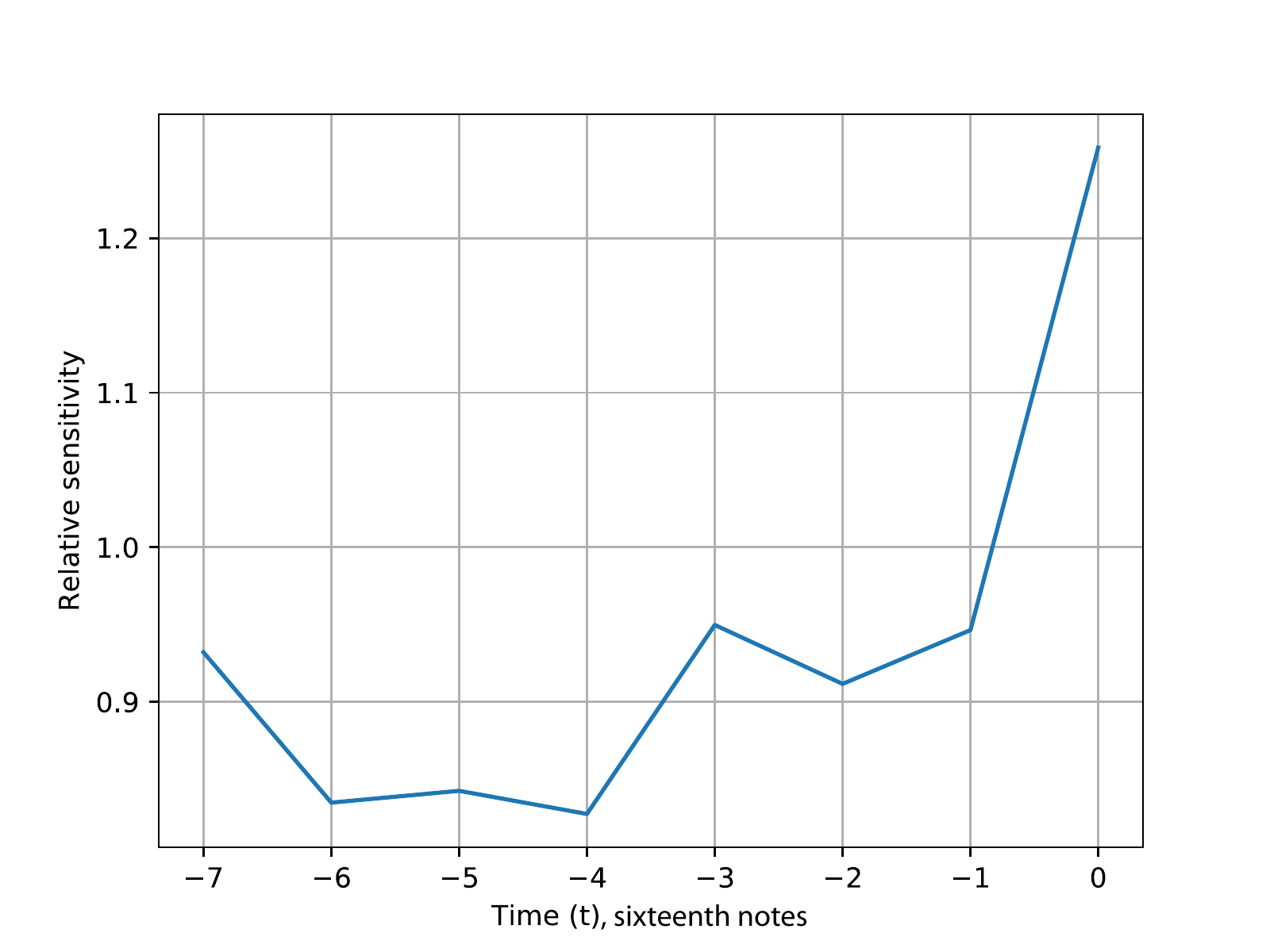}
\caption{Absolute sensitivity of the model when looking backwards on the temporal context, averaged over the whole dataset.}
\label{fig:sensitivity_context}
\end{center}
\end{figure}

The sensitivity of the model to specific context information provides important insights into the functioning of the model.
A common way of determining a networks sensitivity is by calculating the absolute value of the gradients of the networks predictions with respect to the input, holding the network parameters fixed \cite{simonyan2013deep}.
Figure~\ref{fig:sensitivity_context} shows the sensitivity of the model with respect to the temporal context.
The model is particularly sensitive to note occurrences at $t \in \{0,-3,-7\}$.
This shows that the most informative notes for a prediction are direct predecessors ($t=0$), and notes which occur a quarter ($t=-3$) and a half note ($t=-7$, i.e., eight sixteenth notes) before the prediction.

\section{Results and Discussion}\label{sec:res}
The results of the k-nn classification on the raw data and on representations learnt by the model are shown in Table~\ref{tab:knn}.
Classification in the mapping space appreciably outperforms classification in the input space, and obtains similar values for mappings of the original data and the randomly transposed data.
In contrast, when performing classification in the input space the results deteriorate for the randomly transposed input and do not exceed the theoretical lower bound (i.e, always predict all intervals).
As the register and keys of the original data are limited, correlations between absolute and relative pitch exist.
When transposing the input, the classifier cannot make use of these absolute cues for relative pitch any more and performs weakly in the input space.

Figure~\ref{fig:close_matrix} indicates which intervals are close to each other in the mapping space.
An obvious regularity are the slightly brighter k-diagonals (i.e. parallels to the main diagonal) with $k \in \{-24,-12,12,24\}$,
showing that two pitch intervals lead to similar mapping codes when they result in the same pitch class, such as the intervals +8 and -4 semitones, or -7 and -19 semitones.
This is an indication that the model learns the phenomenon of octave equivalence, even if the input to the model represents only absolute pitch.
Another distinct feature is the stripe which is orthogonal to the main diagonal (i.e. where $y=-x$).
This indicates that the model develops some notion of relative distances, by positioning intervals of the same distance (but different signs) close to each other.

Note also that the mappings of certain intervals, notably $6$ and $-6$, are distant to those of most other intervals (dark horizontal and vertical lines).
This likely reflects the fact that tritone intervals are rare in diatonic music, and is further evidence of the musical significance of the learned mappings.

Table~\ref{tab:secdiscovery} shows results of the repeated themes and section discovery task, where the $F_3$ score is a good indicator for the overall performance of the models (see \cite{collins2017discovery} for a thorough explanation on the respective measures).
For the audio data, the current state-of-the-art $F_3$ score was raised from $50.60$ to $51.61$ by our proposed method.
The method performs slightly worse on the symbolic data, which is counterintuitive at first sight, given that results of other models suggest that this task is easier.
Our hypothesis is that for discovery of repeated sections, approximate matching leads to better results than exact comparison, simply because musical variation goes beyond chromatic transposition (towards which our model is invariant).
For approximate matching, a spectrogram representation is better suited than symbolic vectors, as notes are blurred over more than one frequency bin, and harmonics may provide additional cues for a similarity estimation.
The proposed approach is computationally efficient, because the diagonal detector (cf. Equations~\ref{eq:diag1} and~\ref{eq:diag2}) is rather simple and the transposition-invariance of the representations does not require explicit comparison of mutually transposed musical textures.

\section{Conclusion and future work}\label{sec:concl-future-work}
In this paper we have presented a computational approach to deriving (pitch) transposition-invariant vector space representations of music both in the symbolic and the audio domain.
The representations encode pitch intervals that occur in the music in a musically meaningful way,
with tritone intervals---a rare interval in diatonic music---leading to more distinct representations, and octaves leading to more similar representations.
Furthermore, the temporal sensitivity of the model reveals a beat pattern that shows increased sensitivity to pitch intervals occurring at beat multiples of each other.

The transposition-invariance of the representations \linebreak makes it possible to detect transposed repetitions of musical sections in the symbolic and in the spectral domain of audio.
We have demonstrated that this is beneficial in tasks such as the MIREX task \emph{Discovery of Repeated Themes and Sections}.
A simple diagonal finding approach on a transposition-invariant self-similarity matrix produced by our model is sufficient to outperform the state of the art in the audio version of the task.


We believe it is worthwhile to further explore the utility of transposition-invariant music representations for other applications, including speech recognition, music summarization, music classification, transposition-invariant music alignment (including a cappella voices with pitch drift), query by humming, fast melody-based retrieval in large audio collections, and music generation.
First results show that the proposed representations are useful for audio-to-score alignment \cite{arzt2018alignment} and for music prediction tasks \cite{lattner2018predictive}.



\section{Acknowledgments}
This research was supported by the EU FP7 (project \linebreak Lrn2Cre8, FET grant number 610859), and the European Research Council (project CON ESPRESSIONE, ERC grant number 670035). We thank Oriol Nieto for providing us with the source code of his experiments \cite{nieto2014identifying}.



\bibliographystyle{named}
\bibliography{bib/bib_mg,bib/bib_cc,bib/bib_sl}

\end{document}